\def \yskip{\penalty-50\vskip3pt plus 3pt minus 2pt}
\def \reference{\par \yskip \noindent \hangindent .4in \hangafter 1}
\def \abc#1#2#3#4 {\reference#1, {\sl#2}, {\bf#3}, #4}
\def \blank {\lower 5pt\hbox to 0.75in{\hrulefill}}

\def \s{~\rm{s}}
\def \km{~\rm{km}}

\def \K{~\rm{K}}

\def \yrs{~\rm{yrs}}
\def \yr{~\rm{yr}}

\def \lesssim{\mathrel{<\kern-1.0em\lower0.9ex\hbox{$\sim$}}}
\def \gtrsim{\mathrel{>\kern-1.0em\lower0.9ex\hbox{$\sim$}}}

\documentstyle[11pt,aasms,tighten,flushrt]{article}
\begin{document}
%\normalsize
\small

\setcounter{page}{1}
% simwin7.tex (12 Oct. 1999: WITH CORRECTIONS INSIDE, Sub. To ApJ)

\begin{center}
\bf
WHY EVERY BIPOLAR PLANETARY NEBULA IS `UNIQUE'
\end{center}
%\vspace*{2.0cm}

\begin{center}
Noam Soker\\
Department of Physics, University of Haifa at Oranim\\
Oranim, Tivon 36006, ISRAEL \\
soker@physics.technion.ac.il 
\end{center}

%\clearpage
%$$
%$$

\begin{center}
\bf ABSTRACT
\end{center}

 We present the many evolutionary routes that progenitors of
bipolar planetary nebulae (BPNe) can take. Overall, there are
about a hundred qualitatively different evolutionary routes,
hence about a hundred qualitatively different types of BPNe.
Within each type there are quantitative differences as well.
Adding the dependence of the appearance on inclination,
we find that the number of different apparent structures
of BPN is about equal to, or even larger than the number of
known BPNe and proto-BPNe. Accordingly we argue that every BPN is
a `unique' object in its appearance, but all can be explained
within the binary model paradigm.  Therefore, we request a stop
to the attaching of adjectives such as `unique', `peculiar', and
`unusual' to BPNe and proto-BPNe, thereby removing the
need to invoke a new model for almost every `unusual' BPN.
As a case study we try to build a binary model for the proto-BPN
OH231.8+4.2. In our preferred model the AGB Mira-type star has a
main sequence companion of mass $\sim 1 M_\odot$, orbital period of
$\sim 5 \yr$, and eccentricity of $\gtrsim 0.1$. 
 

%\clearpage

% ======================================================================
\section{INTRODUCTION}
% ======================================================================

 With more sensitive observations in recent years a
trend has arisen of attaching adjectives such as `peculiar'
(e.g., Lopez {\it et al.} 2000 to the PN KjPn 8) and `unique'
(e.g., Bourke {\it et al.} 2000 to the Frosty Leo Nebula [IAS 09371+1212])
to bipolar planetary nebulae (BPNe) and proto-BPNe.
 In many cases such adjectives are followed by a claim that a new
theory is required to explain the formation of such `peculiar'
and `unique' BPNe (e.g., Trammell 2000 for the PN AFGL 618).
 BPNe are defined as axially symmetric PNe having two lobes with
an equatorial waist between them. In the present paper BPNe stand both
for BPNe and proto-BPNe.
 The goal is simply to show that in the
binary model for the formation of BPNe there are more than
a hundred different evolutionary routes to form BPNe,
hence every BPN is `unique', and there is no need to invoke new
evolutionary paradigm for each one.

 The observations that other types of binary systems can form bipolar
structures and/or blow jets set the binary model for the formation of
BPNe on solid ground.
 Such systems are symbiotic nebulae (e.g., Morris 1990;
Corradi \& Schwarz 1995; Corradi {\it et al.} 2000),
supersoft X-ray sources (Southwell {\it et al.} 1996;
Southwell, Livio, \& Pringle 1997; Becker {\it et al.} 1998;
Motch 1998; Lee \& Park 1999) and the central binary system of the BPN
NGC 2346 (Bond \& Livio 1990).
 On the theoretical side, Soker \& Rappaport (2000; hereafter SR00)  
show by performing population synthesis simulations that binary systems
can account for the fraction of BPNe among all PNe.
 By achieving its goal, the present paper will strengthen the binary
model for the formation of BPNe. 

 Before listing the different evolutionary routes in section 3, we
take the proto-BPN OH 231.8+4.2 (hereafter OH231; also termed OH 0739$-$14)
as a case study.
This is a well studied proto-BPN (e.g., Cohen {\it et al.} 1985;
Kastner {\it et al.} 1992, 1998; Kastner \& Weintraub 1995; 
Alcolea {\it et al.} 2001, hereafter ABSNZ;
Bujarrabal {\it et al.} 2001, hereafter BCAS).
 ABSNZ raised to question all existing models for the formation
of OH231.
 In particular, ABSNZ argue that it is hard to explain the large
momentum of the outflowing gas in the lobes by existing theories.
 BCAS, in an excellent thorough study of proto-BPNe, extend 
the problem of the momentum source to other objects.
 Disagreeing with the view of ABSNZ and BCAS that existing theories
cannot account for the large momentum in the lobes, I argue in
section 2 that binary models can naturally account for the large
momentum fluxes observed in BPNe (and proto-BPNe), in addition
to the natural explanation of the bipolar structure.
 Cohen {\it et al.} (1985) already suggested that the bipolar structure
is due to a binary companion to the Mira variable, which they
claimed to have detected, and they further pointed to the connection of
this system to symbiotic stars.
 Section 4 contains a short summary.  

% ======================================================================
\section{A CASE STUDY: OH231.8+4.2}
% ======================================================================
% ======================================================================
\subsection{Observed Properties of OH231}
% ======================================================================
 The purpose of this section is to show that the proto-BPN OH231 can
be naturally explained by a binary model, as was suggested already
by  Cohen {\it et al.} (1985), despite the claims of ABSNZ and BCAS
that there is no satisfactory theory to account for this and similar
BPNe.
 Some of the scenarios studied here will be used in the next section
for building a general scheme of models.

 To build a binary model for OH231 we use the following properties
(ABSNZ; BCAS).
(1) A mira variable is the mass-losing star (primary), with the
companion luminosity limited to be $L_2 < 10^3 L_\odot$.
 The limit is deduced from the `nice' Mira-type pulsations seen
in reflection (Kastner {\it et al.} 1992; J. Kastner 2001,
private communication).
(2) The equatorial mass in the bipolar nebula is not much larger than
the mass in the lobes.
 This basically implies that the bipolar flow cannot be
 formed through confinement by a dense equatorial matter.
(3) The sum of the absolute values of the momentum in the lobes
is $p \simeq 27 M_\odot \km \s^{-1}$.
(4) Total kinetic energy of the expanding nebula is
 $E_k \simeq 1700 M_\odot \km^{-1} \s^{-1}$.
(5) Expansion velocities of up to $430 \km \s^{-1}$ are measured.
(6) The nebula contains no, or a very small mass of, photoionized gas.
(7) An increase  by a factor of $\sim 100$ occurred in the mass-loss
rate $\sim 4500 \yr$ ago.
(8) The high momentum flow started $\sim 800 \yr$ ago. 
(9) The acceleration phase of the high momentum flow lasted a short time
   (relative to its $800 \yr$ of age). 
(10) Presently the mass-loss rate by the AGB star is
$\sim 2 \times 10^{-4} M_\odot \yr^{-1}$. 
(11) The lobes are not inflated. 
 There are two basic interactions types between fast and slow
flows in PNe: adiabatic (also termed energy conserving) and
radiative (momentum conserving).
 In the adiabatic case the shocked fast wind cooling time is
much longer than the flow time, while in the radiative
case energy is lost via radiation by the cooling shocked gas. 
 BCAS ruled out the adiabatic case for OH231, mainly on the
ground of the shape of the lobes, which are not
inflated as is expected in the adiabatic flow.

% ======================================================================
\subsection{Models with a WD Companion}
% ======================================================================
% ======================================================================
\subsubsection{A compact WD companion}
% ======================================================================
 An adiabatic flow requires low density and fast CFW (collimated fast wind
blown by the companion), most likely a CFW blown by a WD companion at
 velocities of $v_c \simeq 5000 \km \s^{-1}$.
 To supply most of the kinetic energy cited above in $\tau=800 \yr$
the mass-loss rate from the WD companion into the CFW should be
\begin{eqnarray}
\dot M_c = 2 \times 10^{-7}
\left(\frac {E_k}{10^3 M_\odot \km^{2} \s^{-2}} \right)
\left(\frac {\tau}{500 \yr} \right)^{-1}
\left(\frac {v_c}{5000 \km \s^{-1}} \right)^{-2} M_\odot \yr^{-1}.
\end{eqnarray}
  However, there is a problem with this idea.
 To blow a wind at such a rate the WD must accrete at a rate
several times higher, $\dot M_{\rm WD} \sim 10^{-6} M_\odot \yr^{-1}$.
 A WD accreting at such a rate has a steady nuclear burning
(e.g., Fujimoto 1982) and swells to a radius of
$R_{\rm WD} \gtrsim 0.1 R_\odot$ (Hachisu, Kato, \& Nomoto 1999).
 The escape velocity from the extended wide dwarf surface decreases
to $2300 \km \s^{-1}$ for a $1.3 M_\odot$ WD, with much lower
values for lighter WDs.
 This reduces the CFW speed, hence increases the required mass-loss rate
by a factor of $>4$ compared with that given in equation (1).
 This implies even a higher accretion rate and a larger radius of the
swollen WD, with escape and CFW velocities of
$v_c \lesssim 1000 \km \s^{-1}$.
 From equation (1) the required mass-loss rate into the CFW for
this lower speed is $\sim 10^{-5} M_\odot \yr^{-1}$. 
 At such high accretion rates the WD has swollen to a radius of
$\gtrsim 0.5R_\odot$.
 The problem with this model is the elongated shape of the nebula,
which means there is no massive gas to confine the pressure-accelerated
gas to the polar direction (BCAS).
 This reduces the efficiency of the energy momentum interaction, such
that a much higher mass-loss rate into the CFW is required.
In such a case we are in the regime of momentum-conserving interaction.
 
% ======================================================================
\subsubsection{A Swollen WD companion}
% ======================================================================
 We now examine the case of a CFW from a swollen WD,
such that the interaction is momentum conserving.
 The required momentum constrains the mass-loss rate to 
\begin{eqnarray}
\dot M_c \gtrsim 10^{-4}
\left(\frac {p}{25 M_\odot \km \s^{-1}} \right)
\left(\frac {\tau}{500 \yr} \right)^{-1}
\left(\frac {v_c}{500 \km \s^{-1}} \right)^{-1} M_\odot \yr^{-1}.
\end{eqnarray}
 We note that at very high accretion rates a WD companion will
not ionize the nebula, since it has an extended photosphere
with effective temperature of $\lesssim 3\times 10^4 \K$.
 The required accretion rate for that to occur is 
$\dot M_{\rm WD} = 6 \times 10^{-6} M_\odot \yr^{-1}$ for 
$M_{\rm WD}=0.6 M_\odot$ and
$\dot M_{\rm WD} = 4 \times 10^{-5} M_\odot \yr^{-1}$ for 
$M_{\rm WD}=1.3 M_\odot$ (Hachisu {\it et al.} 1999).
 Therefore, the WD in the momentum-conserving case does not ionize
the circumbinary material (unless the WD is extremely massive and
the accretion rate is fine-tuned).

 The luminosity of an accreting WD due to nuclear burning is
$L>10^4 \K$.
 There are no observational indications for such a luminous
companion. 
 Therefore, in order for this model to work the mass accretion
rate has to be below the steady nuclear burning rate.
 This rate is $10^{-7} M_\odot \yr^{-1}$ for a $1 M_\odot$ WD,
increasing with WD mass (Fujimoto 1982).
 Too massive WD means recurrent outbursts on short time
scales (Fujimoto 1982), and for too light WDs the mass accretion
rate should be even smaller.
 Under what circumstances can the accretion rate from an AGB star
drop by $\sim 3$ orders of magnitude in $\lesssim 300$ years?
There are two possibilities.
 The first is after a Helium flash (thermal pulse) and the second is
a termination of a Roche lobe over flow (RLOF). 
 The mass-loss rate by an AGB star increases somewhat for
a short time after the thermal pulse, and then decreases by two
orders of magnitude (Bl\"ocker 1995).
 The decease in stellar radius after the thermal pulse may lead to
a higher wind velocity, further reducing the accretion rate.
 The RLOF process is unstable in WD-AGB binary systems since the
AGB star at the beginning of the mass-loss process is more
massive than the WD; this case may be applicable to a main
sequence companion, as discussed below. 
  We conclude that a swollen-WD companion model can explain
the properties of OH231, but only if the WD mass accretion rate
is fine-tuned, such that it drops by $\gtrsim 3$ orders of magnitude
in $\lesssim 300$ years, so presently the WD is back to its
normal size and has no nuclear burning on its surface. 
 This is in contradiction with the finding of ABSNZ that the high
mass-loss rate typical of late AGB phases is still going on.
 Another way to ``hide" a WD, swollen or not, is inside
the AGB envelope, i.e., a common envelope system.
 However, it is questionable whether an AGB star harboring a
WD inside its envelope, or a WD that has collided with its core,
will behave like a normal pulsating AGB star, as OH231 does. 
 
 Although unlikely to be the case in OH231, the  process in which a
close WD companion swells to a radius of $\sim 0.1-10 R_\odot$ by
accreting at a very high rate from an AGB star can form very
interesting BPNe.
 When the mass-loss rate by the primary decreases toward the
post-AGB phase in these systems, the WD shrinks and its
temperature rises to the point where it starts ionizing the
nebula, forming a symbiotic nebula on its way to becoming a PNe.

% ======================================================================
\subsection{A Main Sequence Companion}
% ======================================================================
 We are left with a CFW blown by an accreting main sequence companion,
for which the CFW speed is $\sim 500 \km \s^{-1}$ and the 
momentum-conserving case applies.
 The required momentum constrains the mass-loss rate according to
equation (2). 
 To blow such a wind, the main sequence star, which unlike a WD
has no surface nuclear-burning energy source, must accrete at a
rate higher by a factor of $5-10$ than its mass-loss rate to the CFW.
 The total mass blown to the nebula of OH231 at a speed of
$v_c=500 \km \s^{-1}$ is $M_c=0.05 (p/25 M_\odot \km \s^{-1}) M_\odot$.
 During that period the companion has accreted a mass of
$0.25-0.5 M_\odot$ at an average accretion rate of
$\dot M_2 = 0.5-1\times10^{-3} M_\odot \yr^{-1}$.
 However, it seems that the acceleration phase lasted a short
time (ABSNZ), hence the accretion rate was higher. 
 If lasting 250 years, the mass accretion rate is
$\dot M_2 \simeq 1-2\times 10^{-3} M_\odot \yr^{-1}$, which requires
the system to go through a RLOF.
 From the calculations of Prialnik \& Livio (1985) it seems that
a companion of $M_2 \sim 1 M_\odot$ with its outer convective envelope
can accrete such a mass in $\sim 250 \yrs$, even if not thermally relaxed.
 Although Prialnik \& Livio (1985) study the accretion onto 
a fully convective $0.2 M_\odot$ main sequence star, their
calculations show that stars can accrete at very high rates.

  The accretion (from a disk) luminosity is
\begin{eqnarray}
L_{\rm acc} = 3 \times 10^4
\left(\frac {M_2}{ M_\odot} \right)
\left(\frac {\dot M_2}{2 \times 10^{-3} M_\odot \yr^{-1}} \right)
\left(\frac {R}{R_\odot} \right)^{-1} L_\odot.
\end{eqnarray}
 Some fraction of this energy goes to accelerate the CFW, so
the radiative luminosity will be somewhat lower.
 There is no accurate upper limit on the luminosity of
a possible companion, but it is $L_2 <1000 L_\odot$.
 So the present luminosity must be lower than that during the high
mass transfer rate phase. 
  There are two possibilities.
\newline
{\bf (1) A decrease in mass-accretion rate.} 
 For a limit of $L_2 < 1000 L_\odot$ on the companion's
luminosity, the present mass-accretion rate is
$\dot M_2 \lesssim 10^{-4} M_\odot \yr^{-1}$.
 Since the central AGB star loses mass at a rate of
$\sim 2 \times 10^{-4} M_\odot \yr^{-1}$ (ABSNZ), the companion
accretes only a fraction of that, and the limit is met.
 In this case the companion stays outside the AGB envelope.
At the beginning of the mass transfer the stars had about  
equal masses, and the RLOF requires the orbital separation to be
about twice the AGB radius, hence $a\simeq 4 AU$, and the
orbital period is $\sim 5 \yrs$.
  The parameters of such a system are similar in many respects to
those of the Red Rectangle, with one significant difference:
 in the proposed model for OH231 a large amount of gas was
transferred from the AGB star to its companion, part of which was
ejected in the CFW, while in the Red Rectangle only a small amount of
gas was transferred.
 Namely, either the Red Rectangle avoided a RLOF
(Waelkens {\it et al.} 1996), or else the RLOF occurs only after a
substantial part of the AGB envelope has been removed (Van Winckel 2001).
 There are relatively many post-AGB binary systems with similar
parameters to that of the Red Rectangle (Van Winckel 1999),
in particular HD213985, which has a companion with a mass of
$M_2 \simeq 2 M_\odot$ (Van Winckel 1999).
 The companion has probably gained its mass by accretion
(H. Van Winckel 2001, private communication).
 Like these systems, it is quite possible that the large mass transferred
between the stars in OH231 increased the eccentricity (Soker 2000).
 Therefore, if this is the correct scenario for OH231, we predict
that the central system is a binary system where there is a $1 M_\odot$
main sequence companion to the AGB star, at an orbital period of 
$\sim 5$ years, and eccentricity of $e>0.1$. 
\newline
{\bf (2) Common envelope.} 
 Another way by which a companion can escape detection is a 
common envelope evolution.
 Since the AGB star shows regular pulsational behavior
 (Kastner {\it et al.} 1992) the companion
can't be too massive, $M_2 < 0.5 M_\odot$.
 The companion can exists inside the envelope, or it has collided
with the core of the AGB star. 
 The CFW which supplies most of the momentum of the nebula
was blown during a short time, few$\times 100$ years, before
the companion entered the envelope. 
 It is also plausible that during a collision between a star and
the AGB core a disk is formed with two jets.
  The morphology of the BPN NGC 2346, with its 16-day orbital
period binary (Bond \& Livio 1990), shows that the bipolar morphology
is not due to a collision. 

 To summarize, we propose the following model for the formation
of the bipolar nebula of OH231.
  Due to the increase in the AGB stellar radius, possibly a fast
increase after a thermal pulse, the system entered a strong tidal
interaction phase $\sim 4500 \yr$ ago.
 This increased the mass-loss rate, and caused the orbit to shrink
because of orbital angular momentum loss to the AGB envelope
and wind.
 At the same time the radius of the AGB star further increased
due to its mass-loss (Iben \& Livio 1993) and evolution, and its mass
decreased while the accreting main sequence companion stared to
blow a weak CFW.
 Eventually the AGB star filled its Roche lobe, $\sim 800 \yr$ ago.
 It is quite plausible that the AGB star filled its Roche lobe
only during the maximum radius phase during a pulsation cycle.
 For several hundred years the mass transfer rate was very
high and the accreting main sequence companion blew a strong CFW,
which supplied the momentum of the polar flow observed today.
 Next, either RLOF ended and mass-accretion rate decreased by
2 orders of magnitude while the companion stayed outside the
AGB envelope, or the system went through a common envelope phase.
 In the first case the companion has a mass of $M_2 \gtrsim 1 M_\odot$
and the system is similar to the binary systems at the center of the
Red Rectangle and other similar systems (Van Winckel 1999),
while in the latter case the companion is lighter,
$M_2 \lesssim 0.5 M_\odot$, hence the RLOF is unstable,
and the present system may be similar to the one at the center
of NGC 2346.
 Quantitative detailed study is required here after more
observational constraints are available, e.g., limits on
the luminosity of a companion, a limit or detection of AGB
envelope rotation, and a limit or detection of nonradial pulsation
modes, which may hint at a companion inside the envelope and/or
rotation.

 Finally, the companion to the AGB star can explain the departure of
OH231 from axisymmetry (Soker \& Hadar 2001 and references therein).
 The type of departure from axisymmetry is the `bent' type
according to the classification of Soker \& Hadar (2001),
i.e., the two lobes of OH231 are bent to the same side.
This is most clearly seen in the polarization maps of
Kastner \& Weintraub (1995).

% ======================================================================
\section{EVOLUTIONARY ROUTES}
% ======================================================================

 ABSNZ term OH231 a `unique object'.
Similar terms, e.g., `peculiar object', are used in many papers
discussing BPNe.
 We argue that most BPNe (including proto-PNe) are unique
in the sense that the stellar binary model for the formation of
BPNe has many qualitatively different evolutionary routes.
 The number of different evolutionary routes is about
equal to the number of well resolved BPNe.
Some routes are rare, and no known BPNe belong to them, some routes
are common and they contain several known BPNe, and many routes have
only one `unique' known object belong to them.
 Even within an evolutionary channel, quantitative differences exist,
e.g., the opening angle of the CFW (if narrow they
are termed jets), which may cause BPNe to look `unique'.
 Therefore, our view is that there is nothing unique about
`unique' BPNe in the frame of the stellar binary model for
their formation.
 At present, however, with most bipolar PNe we cannot tell the
routes which led to their formation; for this, gasdynamical
simulations of the interaction between the slow AGB wind and
the CFW blown by the companion are required on the theoretical side,
while detection of companions in the center of bipolar PNe are required
on the observational side.

 The main physical parameters characterizing binary progenitors of
BPNe are (a) orbital semimajor axis; (b) eccentricity; (c)
the initial mass and evolution (e.g., helium flash at the end
of the AGB) of the mass-losing star; (d) mass and type (main sequence
or WD) of the companion.
 These parameters can lead to many qualitatively different types of
evolution during the final stages of the AGB and post-AGB phases
(and of course infinite quantitatively different evolutionary paths).
 The different evolutionary types are summarized in Table 1. 
The basic assumption is that most BPNe are formed by a CFW
(collimated fast wind) blown by an accreting companion
(Morris 1987; SR00).
 The basic processes that determine the type of BPNe, which
are summarized in Table 1, are:
\newline
{\it (1) CFW type.} The CFW (or jets) blown by the accreting
companion can be either strong or weak relative to the slow
wind blown by the mass-losing star, as defined in SR00.
In eccentric orbits the CFW may be strong near periastron
(when accretion rate is high) and weak (or not exist
at all) near apastron (Soker 2001a). % wbinary2.tex
\newline
{\it (2) Type of companion and its response to accretion.}
 The companion can be a main sequence star, when the initially more
massive star is the present AGB star, or it can be a WD if the
initially less massive star is the present AGB star (see SR00
for more details). 
 The WD response to accretion depends on its mass and on accretion rate
(e.g., Fujimoto 1982; Hachisu {\it et al.} 1999).
 We distinguish four main cases.
 For low accretion rates, but not too low, so that the WD still blows
a CFW, there is no steady nuclear burning; instead the WD experiences a
symbiotic nova type event, i.e., an outburst
(e.g., Prialnik \& Kovetz 1995).
After the eruption the WD is hot and it enters a supersoft X-ray phase.
 The range of mass accretion rates for this process depends on the WD
mass, but can be approximately taken to be
$10^{-8} \lesssim \dot M_a \lesssim 10^{-7} M_\odot$.
 The descendant PN structure depends on whether or not an outburst
occurs during the final few thousands years.
 For higher accretion rates, $\dot M_a \gtrsim 10^{-7} M_\odot \yr^{-1}$,
there is a steady nuclear burning on the WD surface.
 For lower accretions rates in this range,
$10^{-7} \lesssim \dot M_a \lesssim {\rm few} \times 10^{-6} M_\odot$
(the exact range depends on the WD mass), the WD is a supersoft
X-ray source.
 For very high accretion rates,
$\dot M_a \gtrsim 10^{-5} M_\odot \yr^{-1}$, the WD swells
substantially, up to several solar radii (Hachisu {\it et al.} 1999),
and its surface temperature is $T_{\rm ph} \lesssim 30,000 \K$ so
that it stops ionizing the nebula around it, and the escape velocity
decreases to several hundred $\km \s^{-1}$.
 The structure of the descendant PN depends on whether or not the WD
ionizes the circumstellar medium.
 Of course, a wide range of the WD's wind properties exists within
each of these possibilities, further enlarging the variety
of descendant PN structures.
\newline
{\it (3) Tidal interaction.} Since tidal interaction, as manifested
in the synchronization and circularization times, is very
sensitive to the orbital separation, in most cases there is
either a strong tidal interaction or negligible interaction.
 Strong interaction means for our purposes that there is a
synchronization between the orbital period and the spin period
of the AGB mass-losing star.
 This leads to a high equatorial mass-loss rate. 
 When the two stars are close enough, the systems may enter a RLOF,
or even a common envelope (Iben \& Livio 1993).
\newline
{\it (4) Precession.} The CFW (or jets) blown by the companion
may precess, leading to the formation of a point symmetric PN.

 Some of the evolutionary routes apply to other types of systems,
as indicated in the table.
 The connection of these systems, e.g., symbiotic systems and supersoft
X-ray sources, to PNe were suggested many times in the past.
 We also note that in some cases some of the evolutionary routes
will form elliptical PNe rather than BPNe, e.g., when both
a weak tidal interaction and a weak CFW exist (Soker 2001a).

 The numbers of the different possibilities in each one of these
five parameters are listed in Table 1.
 If the processes were independent, there would be
$3 \times 5 \times 4 \times 2 = 120 $ different
evolutionary routes of BPNe formation. 
However, the different processes listed above are not
completely independent, so the number of routes is much lower.
 For example, a very high accretion rate requires a close orbit,
which in most cases results in strong tidal interaction.
 The connection between the different parameters and the probability
for each route requires a separate study, e.g., a population synthesis.
 On the other hand, several other factors increase the number of
different appearances of bipolar PNe.
(1) In many cases a system will move from one type to another.
For example, if the AGB wind's mass-loss rate increases and/or its
speed decreases in the very last stages of the FIW (superwind) phase,
the accretion rate by the companion increases, and a weak
CFW may become strong. 
 As a result, the descendant PN will contain an outer
structure formed by a weak CFW, and an inner region formed by
a strong CFW.
  The opposite transition occurs when the mass-loss
rate decreases during the proto PN phase.  
(2) Back flow. 
 An accretion by the post-AGB star of material flowing back may
lead to a CFW blown by the post-AGB star (Soker 2001b).
(3) Orientation.
 The inclination angle of the symmetry axis of a BPN (or a proto-BPN)
influences its apparent morphology as well
(e.g., Su, Hrivnak, \& Kwok 2001).
(4) Thermal pulse.
During a thermal pulse (helium flash) the radius and mass-loss rate
of the AGB star increase; a short time later they decrease (Bl\"ocker 1995).
 If a pulse occurs during the FIW it may have an imprint
on the descendant PN.

 Overall, we estimate that binary progenitors of BPNe can
evolve through $\sim 100$ different evolutionary routes.
 This number is about equal to the number of well resolved
BPNe (and proto-BPNe).
 Some evolutionary routes are rare, and contain no known PNe;
some contain several PNe, but considering the quantitative differences
within each evolutionary route, we can safely conclude that
{\it each BPN is a `unique' object} (as stated above, at present we
cannot connect PNe to the exact evolutionary route).
                                           
% ======================================================================
\section{SUMMARY}
% ======================================================================
 In the present work we strengthened the binary model for the
formation of BPNe.
 In that model the companion influences the mass-loss process mainly
during the late AGB phase of the primary mass-loss star, hence the
bipolar structure already exists in the post-AGB phase, and the model
accounts for proto-BPNe as well.
 The main goal was to show that in the binary model there are many
evolutionary routes to form BPNe, hence every BPN is `unique', and
there is no need to invoke new evolutionary paradigm for each BPN,
as some papers have argued in recent years.
 As a case study, we tried to build a model for the formation of
the proto-BPNe OH231.8+4.2, which according to ABSNZ and BCAS
is a challenge to existing theories. We argued in section 2 that
the binary model can naturally account for its properties.
 We predicted that most likely the central AGB star of OH231
 has a main sequence companion of mass $\sim 1 M_\odot$, orbital period of
 $\sim 5 \yr$, and eccentricity of $\gtrsim 0.1$. 
 In section 2 we also discussed several other evolutionary
routes, extending the list for general cases in section 3. 
 The main qualitative parameters which determine the different
evolutionary routes are listed in the first column of Table 1.
 Within each evolutionary route there are quantitative differences,
which further enrich the variety of descendant PNe morphologies.
 We noted that some routes, e.g., weak tidal interaction and weak
CFW (collimated fast wind) will lead mainly to elliptical PNe rather
than BPNe (Soker 2001a).

We request a stop to attaching adjectives such as, `unique',
`peculiar', and `unusual' to BPNe and proto-BPNe, since
almost every BPNe is `unique' in the evolutionary route of
its progenitor binary star.

%====================================================================
%====================================================================
\bigskip

{\bf ACKNOWLEDGMENTS:}
 I thanks the following people for their help:
Javier Alcolea, Valentin Bujarrabal, Joel Kastner, Mario Livio,
Saul Rappaport, and Hans Van Winckel.
This research was supported in part by grants from the
US-Israel Binational Science Foundation.
% ===================================================================

% ===================================================================

\end{document}